\documentclass[12pt]{article}
\usepackage{graphicx}
\usepackage[bookmarksnumbered,colorlinks,plainpages]{hyperref}

\begin{document}

\title{Measuring information growth in fractal phase space}

\author{Q. A. Wang and A. Le M\'ehaut\'e \\
Institut Sup\'erieur des Mat\'eriaux du Mans, \\
44, Avenue F.A. Bartholdi, 72000 Le Mans, France}

\date{}

\maketitle

\begin{abstract}
We look at chaotic systems evolving in fractal phase space. The entropy change in time due to
the fractal geometry is assimilated to the information growth through the scale refinement.
Due to the incompleteness, at any scale, of the information calculation in fractal support,
the incomplete normalization $\sum_ip_i^q=1$ is applied throughout the paper. It is shown that
the information growth is nonadditive and is proportional to the trace-form
$\sum_ip_i-\sum_ip_i^q$ so that it can be connected to several nonadditive entropies. This
information growth can be extremized to give, for non-equilibrium systems, power law
distributions of evolving stationary state which may be called ``maximum entropic evolution''.
\end{abstract}


\vspace{1cm}

\section{Introduction}

The aim of this paper is to study the information evolution of special systems whose phase
space volumes are or map into fractals or multifractals (e.g., dissipative chaotic systems
whose phase space volume maps into fractal attractors for long time $t\rightarrow\infty$).

One of the motivations of this work is related to the generalizations\cite{Esteban} of
Boltzmann and Gibbs entropies which are, in general, connected to dense phase space
($\Gamma$-space). Some of these generalized entropies, e.g., Havrda-Charvat-Tsallis
one\cite{Havrda,Tsal88} or R\'enyi one\cite{Reny66}, are believed to have some connections
with fractal geometry and chaotic behavior of non-equilibrium systems at stationary states.
Our idea in this work is to look at the geometry of the fractal phase space, to calculate the
information from simple geometrical consideration, and to see what is the property of the
evolution of this information and if it has something to do with the generalized entropies.

We suppose an ensemble moving in a fractal phase space, the volume of its initial condition
gradually mapping into fractal structure. That is, as its trajectories run over all the
permitted phase points, its phase volume is covered more and more such that a scale refinement
would be necessary to calculate the occupied volume more and more exactly. Consequently, the
long time behavior of the system of interest is likened to its ensemble behavior : the
information evolution in time is estimated by the information growth during the scale
refinement. This is a crucial point of this work. Due to the incompleteness of the counting of
state points or of the calculation of geometrical elements of fractal structure at any given
scale\cite{Wang04}, the discussion will be made on the basis of the so called incomplete
normalization of probability proposed for the study of the complex systems having physical
states accessible to the systems but inaccessible to theoretical
treatments\cite{Wang04,Wang01,Wang02a,Wang03b}.

\section{Fractal phase space normalization}
To take into account the incompleteness of countable information for a fractal phase space of
dimension $d_f$ embedded in a Euclidean space of dimension $d$, we have put\cite{Wang04} :
\begin{equation}                                            \label{2}
\sum_{i_k=1}^{v_k}p_{i_k}^q=1,
\end{equation}
where $v_k$ is only the number of the accessible states at the $k^{th}$ iteration, $q$ is
given by $q=d_f/d$, and $p_{i_k}=s_{i_k}/S_0$ with $s_{i_k}$ the volume of the $i^{th}$
element of the fractal at the $k^{th}$ iteration\cite{Wang04,Wang03b} and $S_0$ the volume of
the phase space containing the fractal.

The normalization in Eq.(\ref{2}) for incomplete probability distribution was for the first
time discussed by Tribus\cite{Tribus} some decades ago in the context of probability theory
applied to decision theory in engineering and economics where $p_i^q$ was one of the solutions
to the functional equation for logic induction and might be considered as an effective
probability. The probability defined in Eq.(\ref{2}) is different from the usual frequency or
time definition which is normalized. Here the probability $p_i$'s do not sum to one because it
is the ratio of a non-differentiable fractal structure to an integrable and differentiable
homogeneous space. This definition is analogous to that proposed in \cite{Hilb94} to define
probability $p_i$ by the ratio of the number of trajectories (volume $s_i$) on the element
$s_i$ to the total number of trajectories (total volume $S_0$) of the initial conditions.

Eq.(\ref{2}) has been called incomplete normalization. Its incompleteness means that the sum
over all the $v_k$ elements at the $k^{th}$ iteration does not mean the sum over all the
states of the system under consideration, because the volume $s_k(i)$ does not represent the
real number of states or trajectories on the element which, as expected for any fractal
structure, evolves with $k$. So at any given order $k$, the summation over all possible
elements is not equal to the complete summation over all possible states.

\section{Information growth due to fractal}
So there is expansion or contraction of state volume when we refine the phase space
scale. This evolution of the accessible phase volume of a system during the scale
refinement should be interpreted as follows : the extra state points
\begin{equation}                                            \label{3}
\Delta_{i_k}=\sum_{j_{k+1}=1}^{n_{i_k}}s_{j_{k+1}}-s_{i_k}
\end{equation}
acquired from certain element $i_k$ at the iterate of $(k+1)^{th}$ order are just the number
of unaccessible states at $k^{th}$ order with respect to $(k+1)^{th}$ order, where $n_{i_k}$
is the number of elements $s_{j_{k+1}}$ replacing, at $(k+1)^{th}$ iterate, the element
$s_{i_k}$. $\Delta_{i_k}>0$ (or $<0$) means that we have counted less (or more) states at
$k^{th}$ order than the states really accessible to the system. $\Delta_{i_k}$ contains the
{\it accessible information gain} (AIG) through the $(k+1)^{th}$ iterate.

To illustrate the relation between this ``hidden information" and the parameter $q$, let us
suppose that {\it the probability density $\rho$ is scale-invariant} with $\int_{S_0}
\rho^qds=1$ according to incomplete normalization. At the iterate of order $k$, the
information content on $s_k(i)$ is given by
\begin{equation}                                            \label{4}
I_k(i)=\int_{s_{i_k}} I(\rho)ds
\end{equation}
where $I(\rho)$ is the information density. At $k+1$ order, we have
\begin{equation}                                            \label{5}
I_{k+1}(i)=\int_{\sum_{j_{k+1}=1}^{n_{i_k}}s_{j_{k+1}}} I(\rho)ds.
\end{equation}
Now we suppose constant $\rho$ over each element ${i_k}$, i.e., $p_{i_k}\propto s_{i_k}$. This
is true for very small elements after a long evolution of the system of interest or for the
case where every phase point is visited with equal probability. Then AIG reads
\begin{equation}                                            \label{6}
\Delta I_k(i)=I_{k+1}(i)-I_k(i)=\int_{\Delta_{i_k}} I(\rho)ds =I(\rho)\Delta_{i_k}
\end{equation}
The relative AIG is given by
\begin{eqnarray}                                            \label{7}
R_{k\rightarrow (k+1)}(i_k) &=& \Delta I_k(i)/I_k(i) \\ \nonumber  &=&
\sum_{j_{k+1}=1}^{n_{i_k}}\frac{p_{j_{k+1}}}{p_{i_k}}-1.
\end{eqnarray}
Eq.(\ref{7}) suggests that the average value or the expectation of the relative AIG over all
the fractal may be calculated with two possible methods.

\begin{enumerate}

\item By unnormalized expectation :
\begin{eqnarray}                                            \label{7a}
\bar{R}_{k\rightarrow (k+1)}&=&\sum_{i_k=1}^{v_k}p_{i_k}R_{k\rightarrow (k+1)}(i_k)
\\ \nonumber  &=& \sum_{i_{k+1}=1}^{v_{k+1}}p_{i_{k+1}}-\sum_{i_k=1}^{v_k}p_{i_k}
\end{eqnarray}
where $v_k$ (or $v_{k+1}$) is the total number of elements of the fractal at $k^{th}$ (or
$(k+1)^{th}$) iteration.

\item Or by normalized expectation with escort probability :

\begin{eqnarray}                                            \label{7b}
\tilde{R}_{k\rightarrow (k+1)}&=&\frac{\sum_{i_k=1}^{v_k}p_{i_k}R_{k\rightarrow
(k+1)}(i_k)}{\sum_{i_k=1}^{v_k}p_{i_k}}
\\ \nonumber  &=&
\frac{\sum_{i_{k+1}=1}^{v_{k+1}}p_{i_{k+1}}}{\sum_{i_k=1}^{v_k}p_{i_k}}-1.
\end{eqnarray}

\end{enumerate}
The total AIG from $0^{th}$ up to $k^{th}$ iteration is then given by
\begin{eqnarray}                                            \label{8}
R_k=\bar{R}_{0\rightarrow k} =\tilde{R}_{0\rightarrow k}= \sum_{i_k=1}^{v_k}p_{i_k}-1 =
\sum_{i_k=1}^{v_k}(p_{i_k}-p_{i_k}^q)
\end{eqnarray}
because $\sum_{i_0=1}^{v_0}p_{i_0}=S_0/S_0=1$.

So from just geometrical consideration, we end with an information change which is
proportional to the trace-form $\sum_{i_k=1}^{v_k}(p_{i_k}-p_{i_k}^q)$. We see that if $q=1$
or $d=d_f$, the fractal structure does not exist any more, and $R_k=0$. So this is indeed an
information evolution due to the fractal geometry of the phase space.

If one want to write $\wp_i=p_i^q$, then $R_k$ reads $R_k =
\sum_{i_k=1}^{v_k}\wp_{i_k}^{1/q}-1$ with $\sum_{i_k=1}^{v_k}\wp_{i_k}=1$. $\wp_{i_k}$ can be
seen as a complete probability distribution. In this case, do not forget that $\wp_{i_k}$ is
not the real probability $p_{i_k}$.

If we divide certain functional of $R_k$ by the dimension difference $d_f-d$, we can get
several of the trace form entropies in the long list\cite{Esteban} posited, from mathematical
considerations, as possible alternatives to Gibbs or Shannon entropy $S=-\sum_ip_i\ln p_i$.
For example, $\frac{R_{k}}{d_f-d}$ gives the Havrda-Charvat-Tsallis entropy
$S_q=\frac{\sum_ip_i-\sum_i p_i^q}{q-1}$\cite{Havrda,Tsal88}, $\frac{\ln(R_{k}+1)}{d_f-d}$
gives R\'enyi entropy $S^R=\frac{\ln\sum_i \wp_i^{\alpha}}{1-\alpha}$ ($\alpha=1/q$) for
complete distribution\cite{Reny66} or $S^R=\frac{\ln\sum_i p_i}{q-1}$ for incomplete
distribution\cite{Bashkirov}, and $\frac{(R_{k}+1)^{d_f/d}-1}{d_f-d}$ gives Arimoto entropy
$S^A=\frac{(\sum_i\wp_i^{1/q})^q-1}{q-1}$\cite{Arimoto}, among others. It is worth noticing
that all these entropies recover Gibbs-Shannon entropy $S$ when $q\rightarrow 1$ or
$d_f\rightarrow d$.

\section{Some properties of $R_k$}

\begin{enumerate}

\item Nonadditivity : for a fractal of dimension $d_f$ composed of two sub-fractals 1 and
2 of dimension $d_{f_1}$ and $d_{f_2}$ satisfying product joint probability
$p_{{i_{k_1}}{i_{k_2}}}=p_{i_{k_1}}p_{i_{k_2}}$, it is easy to show the following
nonadditivity :
\begin{eqnarray}                                            \label{11}
R_{{k_1}{k_2}}(1+2)=R_{k_1}(1)+R_{k_2}(2)+R_{k_1}(1)R_{k_2}(2).
\end{eqnarray}

\item Concavity and convexity : $R_k$ is positive and concave for $q>1$ and negative and
convex for $q<1$, as shown in Figure 1.

\item For a self-similar fractal whose $n_{i_k}$ is the same for all elements and for any
iteration order, the AIG per iteration given by Eq.(\ref{7}) is independent of $i_k$ :
\begin{eqnarray}                                            \label{13}
R = \sum_{j_{k+1}=1}^{n}\frac{p_{j_{k+1}}}{p_{i_k}}-1 =\sum_{j_{k+1}=1}^{n}r_j-1,
\end{eqnarray}
where $r_j$ is the scale factors of the $j^{th}$ element of the $n$ ones for an iteration. If
all $r_j$'s are equal : $r_j=1/m^d$, we get
\begin{eqnarray}                                            \label{14}
R =n/m^d-1=(1-q)\ln_q(1/m^d)
\end{eqnarray}
where $\ln_q$ is called generalized logarithm and given by $\ln_q
x=\frac{x^{1-q}-1}{1-q}$.

For a Cantor set, $n=2$, $m=3$, $d=1$ et $q=d_f=0.63$, $R=-1/3$.

For a two scale Cantor set with, say, $r_1=1/5$ and $r_2=3/5$, $q$ or $d_f$ is determined by
$(1/5)^q+(3/5)^q=1$, $R=-1/5$.

For von Koch curve, $n=4$, $m=3$, $d=1$, $q=d_f=1.26$, and $R=1/3$.

For Serpinski gasket, $n=3$, $m=2$, and $d=2$, so $d_f=\frac{\ln n}{\ln m}=1.58$,
$q=d_f/d=0.79$ and $R=-0.25$.

For Menger's sponge, $n=20$, $m=3$, and $d=3$, so $d_f=2.73$, $q=d_f/d=0.9$ and $R=-0.26$.

\end{enumerate}

\begin{figure}[ht] \label{f1}
\includegraphics[width=5in,height=4in]{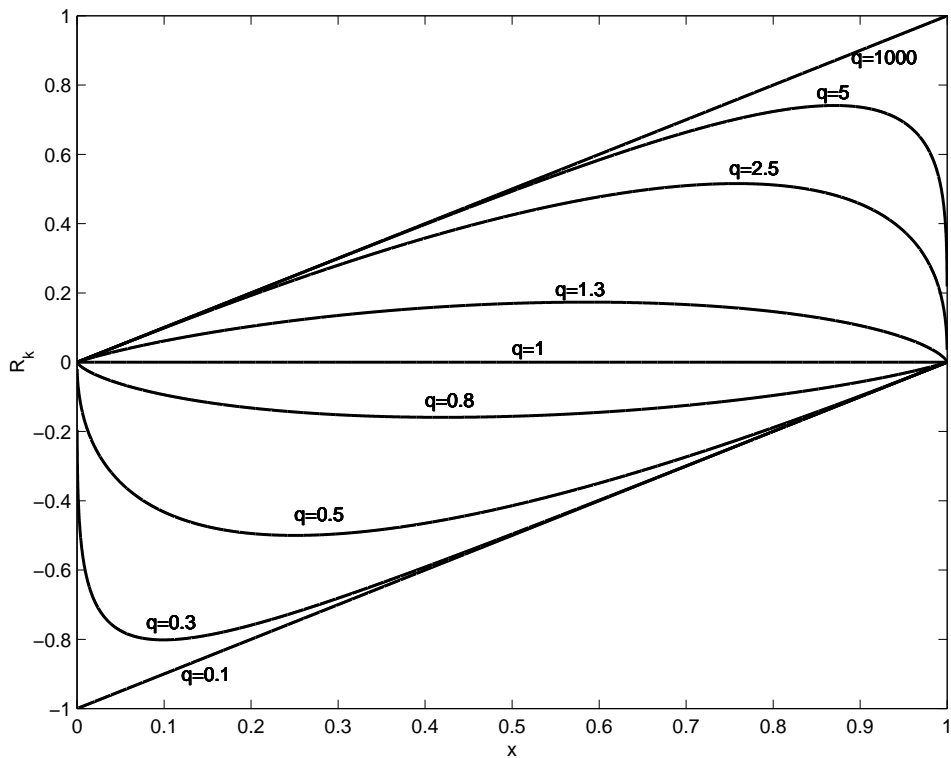}
\caption{The concavity and convexity of the relative information growth $R_k$ for $q>0$ with a
two probability distribution $p_1=x$ and $p_2=(1-x^q)^{1/q}$ satisfying
$\sum_{i=1}^{2}p_i^q=1$. $R_k$ is concave for $q>1$ and convex for $q<1$.}
\end{figure}

\section{Some possible applications}
Now we address stationary probability distribution of non-equilibrium nonadditive systems. The
actual idea of the nonextensive statistical mechanics (NSM) based on Havrda-Charvat-Tsallis
entropy\cite{Tsal88} is to maximize entropy, just for equilibrium state, to get the stationary
power law distributions. This philosophy seems plausible for the systems that have much
smaller relaxation time to reach stationary state than the necessary time interval for the
interaction to change the state. But the method is not justified for stationary state in
general or for non-equilibrium systems which are in evolution with their entropy changing
constantly.

In view of the concavity ($q>1$, $R_k>0$) or convexity ($q<1$, $R_k<0$) of $R_k$ (see Figure
1) and of its connections with different entropies or information measures, it can be easily
checked that the maximum entropy principle is mathematically equivalent to the extremization
of the information growth $R_k$. Our idea here is to extremize $R_k$ in order to obtain the
probability distributions of the stationary states. In this case, the stationary state is a
kind of ``dynamical equilibrium'' or, more precisely, ``equilibrium (stable) evolution'' that,
extremizes the information (entropy) growth $R_k$ and maximizes the entropies at any moment
with respect to other possible evolutions. For this reason, the probability distributions
obtained from extremizing $R_k$ can be called {\it probability distribution for maximum
entropic evolution}. Following are some examples of power law distributions obtained by the
$R_k$-extremum.

\begin{enumerate}

\item {\bf The continued fraction map}\cite{Beck} is given by $x_{n+1}=1/x_n-\lfloor
1/x_n\rceil$ where $\lfloor 1/x_n\rceil$ is the integer part of $1/x_n$ and $x_n$ is a
real number between 0 and 1. The probability distribution is given by $\rho(x)=1/(1+x)\ln
2$ and satisfies $\int_0^1\rho(x)dx=1$. This distribution can be obtained by extremizing
\begin{eqnarray}                                            \label{14a}
R_k=\int_0^1\tilde{\rho}(x)dx-1
\end{eqnarray}
under the constraints associated with the normalization $\int_0^1\tilde{\rho}^q(x)dx=1$ and
with the unnormalized expectation $\overline{x}=\int_{[0<x<1]}\tilde{\rho}(x)xdx$ or with the
normalized expectation $\widetilde{x}=
\frac{\int_{[0<x<1]}\tilde{\rho}(x)xdx}{\int_{[0<x<1]}\tilde{\rho}(x)dx}$, we can obtain the
distribution
\begin{eqnarray}                                            \label{15a}
\rho(x)=\tilde{\rho}^q(x)=\frac{c}{(1-\beta x)^{q/(1-q)}}
\end{eqnarray}
where $q=1/2$, $\beta=-1$ is the Lagrange multiplier associated with $\overline{x}$ and the
constant $c=1/\ln 2$ is determined by the normalization.

\item {\bf The Ulam maps} is a logistic map $x_{n+1}=1-\mu x_n^2$ ($-1<x<1$) with $\mu=2$. Its
probability distribution is given by $\rho(x)=\frac{1}{\pi(1-x^2)^{1/2}}$ which is
normalized\cite{Beck}. By extremizing $R_k$ under the constraints associated with
$\int_{-1}^1\tilde{\rho}(x)^qdx=1$ and with the unnormalized expectation
$\overline{x^2}=\int_{-1}^1\tilde{\rho}(x)x^2dx$ or with the normalized expectation
$\widetilde{x^2}= \frac{\int_{-1}^1\tilde{\rho}(x)x^2dx}{\int_{-1}^1\tilde{\rho}(x)dx}$, we
can obtain a distribution
\begin{eqnarray}                                            \label{15}
\rho(x)=\tilde{\rho}^q(x)=\frac{c}{(1-\beta x^2)^{q/(1-q)}}
\end{eqnarray}
where $q=1/3$, $\beta=1$ is the Lagrange multiplier associated with $\overline{x^2}$ and the
constant $c=1/\pi$ is determined by the normalization.

\item {\bf The Zipf-Mandelbrot's law} $\rho(x)=\frac{A}{(1-Bx)^\gamma}$ is an useful chaotic
distribution for some complex systems\cite{Alm}, where $A$, $B$ and $\gamma$ are constants.
Using the same machinery as above, one obtain :
\begin{eqnarray}                                            \label{16}
\rho(x)=\frac{A}{(1-\beta x)^{q/(1-q)}}
\end{eqnarray}
where $\beta=B$ is the Lagrange multiplier associated with $\overline{x}$, $1/q=1+1/\gamma$
and $A$ should be determined by the condition $\int_{-1}^1\rho(x)dx=1$.

\end{enumerate}

\section{Conclusion}
In summering, we have discussed the accessible information growth (AIG) during long time
evolution of some chaotic systems or through the scale refinement in their fractal phase
space. AIG turns out to take the trace form $\sum_ip_i-\sum_ip_i^q$ so that it can be
connected with several entropies which generalize the Gibbs-Shannon one. It is argued that,
for non-equilibrium systems, it would be more natural to extremize AIG, instead of maximizing
entropy, to get the stationary probability distributions. It turns out that this extremization
is mathematically equivalent to the maximization of the relevant generalized entropies.
Several power law distributions have been obtained from the extremization of AIG.

On the other hand, the AIG of this work has been calculated for chaotic systems whose phase
space is or maps into fractals. So it may be invalid for other non-equilibrium systems that do
not have this property. What is the information growth in that case? How to obtain the
probability distribution for these non-equilibrium systems? Does the method of extremum
information growth applies in general to them? What is the thermodynamics of the ``dynamical
equilibrium'' of this nonadditive systems in constant evolution? All these questions, together
with other fundamental questions about NSM, still remain open.

\end{document}